%% file: main_PRA-L.tex
\documentclass[10pt,a4paper,aps,twocolumn,nobibnotes,pra,showpacs,superscriptaddress,nofootinbib]{revtex4-2}
\usepackage{color,graphicx}
\usepackage{amssymb,amsmath}
\usepackage[utf8]{inputenc}
\usepackage{hyperref}
\usepackage{bm,bbm}
\usepackage{textcomp,gensymb,float}
\usepackage{amsthm,braket}
\usepackage{mathrsfs,mathtools}
\usepackage{enumerate}
\usepackage[varg]{txfonts} 
\usepackage{subfigure}

\DeclareMathOperator{\Tr}{Tr}

\newcommand{\be}{\begin{eqnarray}}
\newcommand{\ee}{\end{eqnarray}}

\newcommand{\mc}{\mathcal}
\newcommand{\ms}{\mathscr}

\newcommand{\mf}{\mathfrak}

\newcommand{\mbb}{\mathbbm}

\newcommand{\tx}{\text}

\newtheorem*{definition}{Definition}

\begin{document}

\title{Theory-Independent Context Incompatibility: Quantification and Experimental Demonstration}

\author{Mariana Storrer\href{https://orcid.org/0009-0002-4696-6620}{\includegraphics[scale=0.05]{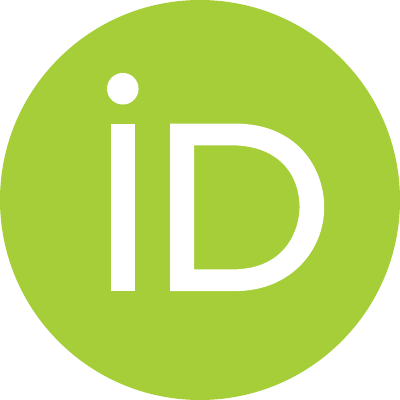}}}
%\email{@fisica.ufpr.br}
\affiliation{Department of Physics, Federal University of Paran\'a, Curitiba, Paran\'a, P.O. Box 19044, 81531-980, Brazil}

\author{Patrick Lima\href{https://orcid.org/0000-0001-9004-141X}{\includegraphics[scale=0.05]{orcidid.pdf}}}
\affiliation{Department of Physics, Federal University of Minas Gerais, 31270-901, Belo Horizonte, MG, Brazil}

\author{Ana C. S. Costa\href{https://orcid.org/0000-0002-4014-0695}{\includegraphics[scale=0.05]{orcidid.pdf}}}
%\email{ana@fisica.ufpr.br}
\affiliation{Department of Physics, Federal University of Paran\'a, Curitiba, Paran\'a, P.O. Box 19044, 81531-980, Brazil}

\author{Sebastião Pádua\href{https://orcid.org/0000-0001-9429-1205}{\includegraphics[scale=0.05]{orcidid.pdf}}}
\affiliation{Department of Physics, Federal University of Minas Gerais, 31270-901, Belo Horizonte, MG, Brazil}

\author{Renato M. Angelo\href{https://orcid.org/0000-0002-7832-9821}{\includegraphics[scale=0.05]{orcidid.pdf}}}
%\email{renato.angelo@ufpr.br}
\affiliation{Department of Physics, Federal University of Paran\'a, Curitiba, Paran\'a, P.O. Box 19044, 81531-980, Brazil}

\date{\today}

\begin{abstract}
The concept of compatibility originally emerged as a synonym for the commutativity of observables and later evolved into the notion of measurement compatibility. In any case, however, it has remained predominantly algebraic in nature, tied to the formalism of quantum mechanics. Recently, still within the quantum domain, the concept of context incompatibility has been proposed as a resource for detecting eavesdropping in quantum communication channels. Here, we propose a significant generalization of this concept by introducing the notion of theory-independent context compatibility, a concept that is trivially satisfied by classical statistical mechanics but is found in conflict with quantum mechanics. Moreover, we propose a figure of merit capable of quantifying the degree of violation of theory-independent context incompatibility, and we experimentally demonstrate, using a quantum optics platform, that quantum systems can exhibit pronounced degrees of violation. Besides yielding a concept that extends to generic probabilistic theories and retrieving the notion of measurement incompatibility in the quantum domain, our results offer a promising perspective on evaluating the role of incompatibility in the manifestation of non-local correlations.
\end{abstract}

\maketitle

%%%%%%%%%%%%%%%%%%%%%%%%%%%%%%%%%%%%%%%%%%%
\textit{Introduction.} Measurement incompatibility lies at the core of quantum foundations and quantum information science~\cite{Heinosaari_2016,rev.jointmeasur}. Initially linked to the uncertainty principle and the noncommutativity of observables~\cite{heisenberg,robertson}, it was later generalized beyond projective measurements to positive-operator valued measurements (POVMs) and joint measurability~\cite{Bush_1986,Heinosaari_2008,Heinosaari_2010}. It underpins fundamental principles such as no-cloning and no-information-without-disturbance and enables tasks ranging from state discrimination to random access codes and nonclassical correlations~\cite{Scarani_2005,Busch_2009,state.disc.carmeli,state.disc.skrz,Frenkel_2015,Wolf_2009,Quintino_2016,Hirsch_2018,Bene_2018,steering,Quintino_2014,Uola_2014,Uola_2015,rev.jointmeasur,borges_14,rev.contextuality,Buscemi_2020}. To probe its role beyond quantum theory and in the classical limit, the notion of \textit{context incompatibility} was introduced and later extended to POVMs~\cite{MSA,Mitra_2021}, capturing state-dependent quantumness and decoherence-induced classical behavior while recovering measurement incompatibility in suitable regimes. However, these formulations remain confined to the quantum formalism, motivating theory-independent generalizations compatible with broader probabilistic frameworks~\cite{cavalcanti,barrett-pqm,Dariano_2017}.

In search of a broader notion of incompatibility—one that captures the classical limit, accommodates generalized measurements, applies to generic probabilistic theories, and recovers measurement incompatibility—we introduce in this Letter the concept of \textit{theory-independent context incompatibility} (TICI). We also propose a measure to quantify violations of theory-independent context compatibility. Using projective measurements on a quantum optics platform, we test our framework through a three-step experiment: (i) entangled photon pairs are generated via spontaneous parametric down-conversion (SPDC); (ii) a general single-photon qubit mixed state is prepared by detecting the partner photon without resolving its polarization; and (iii) sequential measurements are performed on the remaining photon. Our results confirm that nature violates theory-independent context compatibility, in agreement with quantum mechanical predictions. Fundamentally, our work points to an important refinement of the notion of incompatibility underlying Heisenberg’s uncertainty principle and related formulations.

%%%%%%%%%%%%%%%%%%%%%%%%%%%%%%%%%%%%%%%%%%%
\textit{Theory-independent context compatibility.}
We consider a context $\mbb{C}=\{\ms{E},\ms{X},\ms{Y}\}$ composed of generalized measurements $\ms{X}$ and $\ms{Y}$ with respective outcomes $\{x_i\}_{i=1}^d$ and $\{y_j\}_{j=1}^d$, and a preparation state $\ms{E}$. For simplicity, we assume that both measurements have the same number $d$ of outcomes. Let $p_\ms{E}(y_j|x_i)$ denote the probability of obtaining outcome $y_j$ from $\ms{Y}$ given that outcome $x_i$ was previously obtained from $\ms{X}$, with $\ms{E}$ as the initial state. When the outcome $x_i$ is not recorded, the probability of obtaining $y_j$ from $\ms{Y}$ after a nonselective measurement of $\ms{X}$ follows from weighting $p_\ms{E}(y_j|x_i)$ by the occurrence probabilities $p_\ms{E}(x_i)$:
\begin{align}\label{M_X}
p_{\ms{M_X}(\ms{E})}(y_j)\coloneqq\sum_{i=1}^d p_\ms{E}(y_j|x_i)\,p_\ms{E}(x_i). 
\end{align}
This expression defines the generalized nonselective measurement map $\ms{M}_\ms{X}(\ms{E})$. We are now ready to formalize theory-independent context compatibility.

%------------------
\begin{definition}
A context $\mbb{C}=\{\ms{E},\ms{X},\ms{Y}\}$ is said to be compatible if nonselective measurements of $\ms{X}$ and $\ms{Y}$ leave each other's outcome statistics invariant, that is,
\begin{align}\label{eqsdef}
p_\ms{E}(x_i)=p_{\ms{M}_\ms{Y}(\ms{E})}(x_i),\qquad
p_\ms{E}(y_j)=p_{\ms{M}_\ms{X}(\ms{E})}(y_j).   
\end{align}
\label{def}
\end{definition}
%------------------

\begin{figure}[t]
\centering  \includegraphics[scale=0.17]{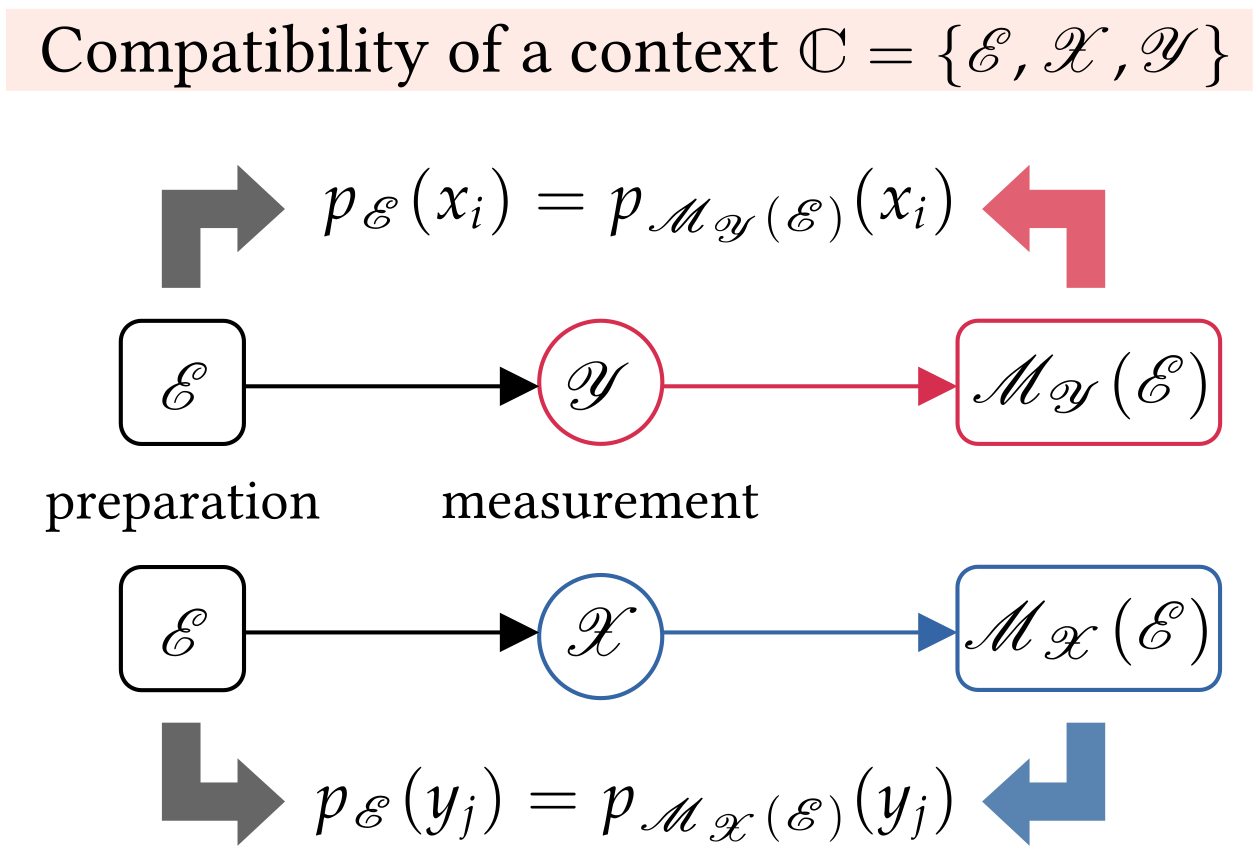}
\caption{Schematic representation of context compatibility. A context $\mbb{C}=\{\ms{E},\ms{X},\ms{Y}\}$ consists of a preparation $\ms{E}$ and generalized measurements $\ms{X}$ and $\ms{Y}$ with outcomes $x_i$ and $y_j$.  Each measurement induces a nonselective map on the preparation, and the context is compatible when the resulting distributions of all $x_i$ and $y_j$ match those obtained directly from $\ms{E}$.}
\label{fig:ContInc_concept}
\end{figure}

This definition is illustrated in Fig.~\ref{fig:ContInc_concept}. It connects to Fine's consistency criterion~\cite{Fine_1982} and is trivially satisfied in classical statistical mechanics (see Supplemental Material~I for details).
%Appendix~A for details).

%%%%%%%%%%%%%%%%%%%%%%%%%%%%%%%%%%%%%%%%%%%%%%%%
\textit{Context compatibility in quantum mechanics.}
Quantum theory does not generally admit joint probability distributions for arbitrary observables (e.g., position and momentum), so violations of criterion~\eqref{eqsdef} are expected. A quantum context $\mbb{C}=\{\rho,A,B\}$ consists of a state $\rho$ and POVMs $A=\{\alpha_i\}$ and $B=\{\beta_j\}$ on a finite-dimensional Hilbert space $\mc{H}$. Using Kraus decompositions $\alpha_i=A_i^\dagger A_i$ and $\beta_j=B_j^\dagger B_j$, the nonselective measurement of $A$ is given by
\begin{align}
\label{eq:nsmm}
\sum_ip_i\tilde{\rho}_i=\sum_iA_i\,\rho\, A_i^\dag\eqqcolon\Phi_A(\rho),
\end{align}
where $\tilde{\rho}_i=A_i\rho A_i^\dag/p_i$ and $p_i=\Tr(\alpha_i\rho)$, with an analogous expression for $\Phi_B(\rho)$. 
In this framework, condition~\eqref{eqsdef} reads 
$\Tr(\alpha_i\rho)=\Tr[\alpha_i\Phi_B(\rho)]$ and 
$\Tr(\beta_j\rho)=\Tr[\beta_j\Phi_A(\rho)]$. 
Since our goal is to show that quantum mechanics violates compatibility, we may restrict to projective measurements. 
With $A_i=\ket{a_i}\!\bra{a_i}$ and $B_j=\ket{b_j}\!\bra{b_j}$, we obtain 
$\Phi_A(\rho)=\sum_i p(a_i)A_i$ and 
$\Phi_{BA}(\rho)\equiv\Phi_B(\Phi_A(\rho))=\sum_{i,j}p(b_j|a_i)p(a_i)B_j$, 
where $p(a_i|b_j)=p(b_j|a_i)=\Tr(A_iB_j)$,
$p(a_i)=\Tr(A_i\rho)$, and $p(b_j)=\Tr(B_j\rho)$. 
Multiplying Eqs.~\eqref{eqsdef} by $A_i$ and $B_j$ and summing over $i$ and $j$ yields:
\begin{align}\label{criteriaq}
\Phi_{A}(\rho)=\Phi_{AB}(\rho)\quad\tx{and}\quad\Phi_B(\rho)=\Phi_{BA}(\rho).
\end{align} 
These equations express context compatibility for projective measurements in quantum mechanics. Violation occurs when $\rho=A_k$ (an eigenstate of $A$) and the basis of $B$ is mutually unbiased with respect to $A$, i.e., $\Tr(A_iB_j)=1/d$.

When restricted to sequential measurements on a single quantum system, the condition in Eq.~\eqref{criteriaq} reduces to the no-signaling-in-time (NSIT) criterion~\cite{Kofler_2013,Uola_2019,Ku_2022}. 
Unlike NSIT, our formulation does not assume Hilbert-space structure or quantum instruments: contexts are defined operationally through preparation and measurement statistics. The approach extends to arbitrary probabilistic theories, treating temporal and spatial scenarios on equal footing, with NSIT arising as a special case of TICI.

%%%%%%%%%%%%%%%%%%%%%%%%%%%%%%%%
\textit{Application: one qubit system.}
To better grasp~\eqref{criteriaq}, we analyze the single-qubit case. In the Bloch representation~\cite{bloch}, we write 
$\rho=\tfrac{1}{2}(\mbb{1}+\vec{r}\cdot\vec{\sigma})$, 
$A=\hat{a}\cdot\vec{\sigma}$, and $B=\hat{b}\cdot\vec{\sigma}$, 
with associated projectors 
$A_i=\tfrac{1}{2}[\mbb{1}+(-1)^i\hat{a}\cdot\vec{\sigma}]$ and 
$B_j=\tfrac{1}{2}[\mbb{1}+(-1)^j\hat{b}\cdot\vec{\sigma}]$, 
where $\hat{a},\hat{b},\vec{r}\in\mathbbm{R}^3$, $||\hat{a}||=||\hat{b}||=1$, and $r=||\vec{r}||\in[0,1]$. 
Direct evaluation gives 
$\Phi_A(\rho)=\tfrac{1}{2}[\mbb{1}+(\vec{r}\cdot\hat{a})(\hat{a}\cdot\vec{\sigma})]$ and 
$\Phi_{AB}(\rho)=\tfrac{1}{2}[\mbb{1}+(\vec{r}\cdot\hat{b})(\hat{a}\cdot\hat{b})(\hat{a}\cdot\vec{\sigma})]$ 
(and analogously for $\Phi_B$ and $\Phi_{BA}$). 
Applying~\eqref{criteriaq} leads to the following vector conditions for compatibility:
\begin{align}\label{angles}
\vec{r}\cdot [\hat{a}-\hat{b}(\hat{a}\cdot  \hat{b})]=0\quad\tx{and}\quad
\vec{r}\cdot[\hat{b}-\hat{a}(\hat{b}\cdot\hat{a})]=0.  
\end{align}
Using the Schatten $2$-norm $||O||_2\coloneqq (\Tr[O^\dag O])^{1/2}$ and defining 
$\vec{u}=\hat{a}-\hat{b}(\hat{a}\cdot\hat{b})$ and 
$\vec{v}=\hat{b}-\hat{a}(\hat{b}\cdot\hat{a})$, 
Eqs.~\eqref{angles} become 
$\mu r\cos\theta_{\vec{r},\vec{u}}=0$ and 
$\mu r\cos\theta_{\vec{r},\vec{v}}=0$, 
where $\mu\equiv||\vec{u}||=||\vec{v}||=\tfrac{1}{\sqrt{8}}||[A,B]||_2$. 
This yields three distinct cases in which the context is compatible:
\begin{enumerate}[(i)]
\item $r=0$, i.e., the maximally mixed state $\rho=\mbb{1}/d$, for which $[A,\rho]=[B,\rho]=0$;
\item $\mu=0$, which holds only when $[A,B]=0$;
\item $\cos\theta_{\vec{r},\vec{u}}=\cos\theta_{\vec{r},\vec{v}}=0$, meaning $\vec{r}$ is orthogonal to the plane spanned by $\hat{a}$ and $\hat{b}$, equivalently $[\rho,[A,B]]=0$.
\end{enumerate}
In the qubit case, these conditions fully characterize compatibility: if none hold, the context is incompatible.

%%%%%%%%%%%%%%%%%%%%%%%%
\textit{Quantification.}
Since the violation of theory-independent context compatibility may depend on system dynamics, it is useful to quantify how far a context deviates from compatibility. 
Using the Kullback–Leibler (KL) divergence~\cite{kullback}, 
$D(P(\ms{X})||P(\ms{Y}))=\sum_i p(x_i)\log_b[p(x_i)/p(y_i)]\ge0$ 
for distributions $P(\ms{X})=\{p(x_i)\}$ and $P(\ms{Y})=\{p(y_j)\}$, 
we define the \textit{theory-independent context incompatibility} (TICI)
\begin{align}
\mathcal{I}_{\mbb{C}}\coloneqq\frac{D\big(P_\ms{E}(\ms{X})|| P_{\ms{M_\ms{Y}(\ms{E})}}(\ms{X})\big)+D\big(P_\ms{E}(\ms{Y})|| P_{\ms{M_\ms{X}(\ms{E})}}(\ms{Y})\big)}{2}
\label{I_C}
\end{align}
of a context $\mbb{C}=\{\ms{E,X,Y}\}$, where $P_\ms{E}(\ms{X})=\{p_\ms{E}(x_i)\}$ (and analogously for the other distributions). 
The logarithm base $b$ is arbitrary. 
By construction, $\mathcal{I}_{\mbb{C}}\ge 0$, with equality if and only if the compatibility condition~\eqref{eqsdef} holds, and it is strictly positive whenever the corresponding distributions differ statistically. 
For a quantum context $\mbb{C}=\{\rho,A,B\}$ with projective measurements, one finds 
$D(P_\rho(A)||P_{\Phi_B(\rho)}(A)) = S(\Phi_A(\rho)||\Phi_{AB}(\rho))$, 
where $P_\rho(A)=\{\Tr(A_i\rho)\}$ and 
$S(\rho||\sigma)=\Tr[\rho(\log_b\rho-\log_b\sigma)]$ is the von Neumann relative entropy~\cite{vonNeumann}. 
It then follows that the incompatibility reads
\begin{align}
\mathcal{I}_{\{\rho,A,B\}}= \frac{S(\Phi_A(\rho)\lVert \Phi_{AB}(\rho))+ S(\Phi_B(\rho)\lVert \Phi_{BA}(\rho))}{2}.
\label{qI_C}
\end{align}
Although Eq.~\eqref{eqsdef} defines compatibility in a fully operational and theory-independent way, its relation to standard notions of measurement incompatibility depends on the underlying theoretical structure. Outside quantum mechanics, concepts such as disturbance or joint measurability are not uniquely defined across generalized probabilistic theories. Within the quantum formalism, however, projective measurements provide a natural operational framework, and the TICI reduces to the state-transformation form in Eq.~\eqref{qI_C}, coinciding with established entropic incompatibility measures.

This raises the question of when context incompatibility reduces to a notion depending solely on the measurements. While an answer for general contexts remains open, Eq.~\eqref{qI_C} already reveals the regimes in which such a reduction occurs in the quantum setting.

For states already compatible with one observable, the incompatibility should reasonably depend only on the other. To examine this, consider the context $\mbb{C}=\{A_k,A,B\}$, where $\rho=\ket{a_k}\!\bra{a_k}$ is an eigenstate of $A=\sum_i a_i\ket{a_i}\!\bra{a_i}$ and $B=\sum_j b_j\ket{b_j}\!\bra{b_j}$. In this case, the quantifier~\eqref{qI_C} reduces to
\begin{align}
\mc{I}_{\{A_k,A,B\}}=-\frac{1}{2}\log_b \left( \sum^d_j |\braket{b_j|a_k}|^4 \right),
\label{teste1}
\end{align}
which is, in fact, a quantifier of measurement incompatibility alone. For mutually unbiased observables, $\mc{I}_{\{A_k,A,B\}}=\tfrac{1}{2}\log_b d$. This value matches the maximum found in numerical simulations of $\mc{I}_\mbb{C}$ over one million randomized qubit contexts ($d=2$). An analogous result holds for $\rho=B_l$.

The TICI quantifier satisfies:
(i) \emph{Positivity and faithfulness}: it is non-negative and vanishes exactly for compatible contexts; 
(ii) \emph{Convexity}: as a KL-based measure, it is convex under classical mixing of preparations; 
(iii) \emph{Monotonicity under coarse-graining}: coarse-graining of outcomes or of the context cannot increase its value; 
(iv) \emph{Reduction to measurement incompatibility}: in the quantum formalism, and for projective measurements on a pure eigenstate, Eq.~\eqref{teste1} shows that TICI reduces to a relative-entropy expression between post-measurement states.

In Ref.~\cite{MSA}, quantum context incompatibility was introduced as an informational resource arising from successive measurements of noncommuting observables. There, incompatibility depends jointly on the state and the measurements and is maximized for mutually unbiased bases. The present framework extends this perspective by (a) providing a fully operational, theory-independent definition of compatibility and (b) introducing a divergence-based quantifier directly extracted from statistics. Thus, TICI serves both as a diagnostic of disturbance and as a resource for temporal protocols in which predictive power and information gain rely on context incompatibility.

\begin{figure}[H]
    \centering  \includegraphics[width=\linewidth]{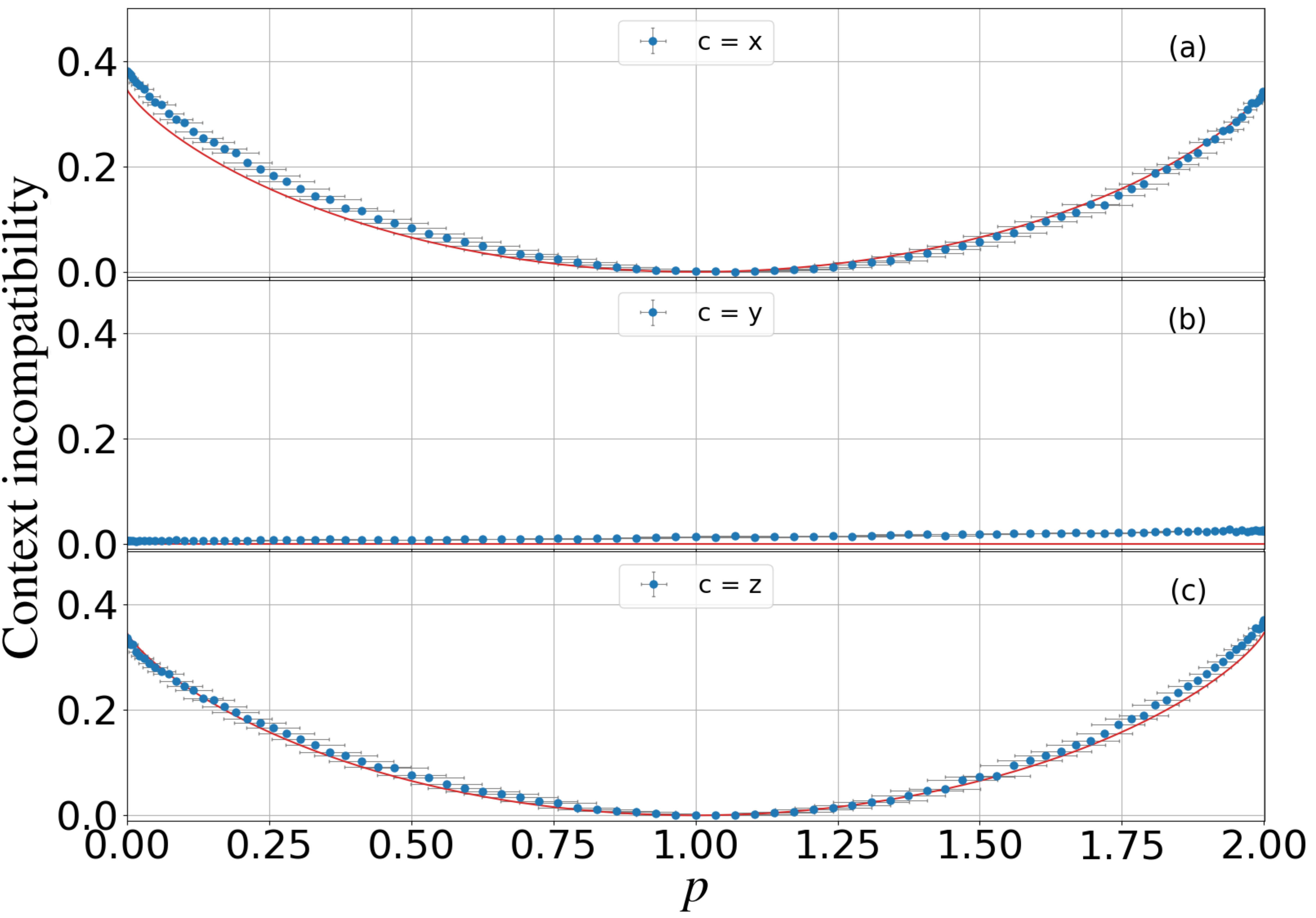}
    \caption{
    Theoretical prediction of $\mathcal{I}_{\mbb{C}}$ (red line) and experimental data (blue dots) for the context
    (a) $\{\rho_x,\sigma_x,\sigma_z\}$, 
    (b) $\{\rho_y,\sigma_x,\sigma_z\}$, and 
    (c) $\{\rho_z,\sigma_x,\sigma_z\}$ as a function of $p$. 
    The $y$-axis error bars are included but not visible (order $10^{-2}$). The calculations were performed using base-$e$ logarithms.
    }
    \label{fig:incompatibility-xyz}
\end{figure}

%%%%%%%%%%%%%%%%%%%%%%%%%%%%%%%%%%%%%%%%%%%%%%%%%%%%%
\textit{Experimental demonstration of context incompatibility.}
We now present experimental evidence that quantum systems violate context compatibility. 
As a testbed, we considered the one-qubit mixed state
$\rho_\mf{c}=\tfrac{p}{2}\mbb{1}+(1-p)\psi_\mf{c}$,
where $\psi_\mf{c}=\ket{\psi_\mf{c}}\!\bra{\psi_\mf{c}}$ with 
$\ket{\psi_\mf{c}}=\cos(\theta/2)\ket{0}+e^{i\phi}\sin(\theta/2)\ket{1}$, 
$p\in[0,2]$, $\theta\in[0,\pi/2]$, and $\phi\in[0,2\pi)$. 
The parameter $p$ interpolates between a pure and a maximally mixed state.
For the measurements, the pure component $\psi_\mf{c}$ was aligned with the $x$-, $y$-, or $z$-axis using the settings 
$\{\theta=\tfrac{\pi}{2},\phi=0\}$, 
$\{\theta=\tfrac{\pi}{2},\phi=\tfrac{\pi}{2}\}$, and 
$\{\theta=0,\phi=0\}$, respectively. 
This yielded three experimental curves of $\mc{I}_{\mbb{C}}$ versus $p$, one for each choice of $\mf{c}$ (Fig.~\ref{fig:incompatibility-xyz}).

To generate the state $\rho_\mf{c}$, we employ a continuous-wave $355\,\mathrm{nm}$ pump laser and two adjacent type-I BiBO (BiB$_3$O$_6$) crystals with orthogonal optic axes~\cite{kwiat_99}, producing polarization-entangled photon pairs via SPDC. The vertically polarized pump first passes through a half-wave plate (HWP$_p$) set at an angle $\theta_p$ relative to its horizontally aligned fast axis, and then through the BiBO crystals, where SPDC generates the entangled trigger-signal pair. 
Symbolically, this process yields

\begin{equation}
    \begin{split}
        \label{eq:spdc}
        E_0\begin{pmatrix}
        0 \\
        1
    \end{pmatrix}
    \xrightarrow{\text{HWP}(\theta_p)}
    E_0\begin{pmatrix}
         \sin{2 \theta_p} \\
         -\cos{2 \theta_p}
    \end{pmatrix} \\ \xrightarrow{\text{SPDC}} \cos{(2 \theta_p)} \ket{HH} + \sin{(2 \theta_p)} e^{i\delta} \ket{VV} = \ket{\Psi} \,,
    \end{split}
\end{equation}
where $|E_0|^2$ is the pump intensity and $\ket{\Psi}$ denotes the two-photon state in the $\ket{HH}$ and $\ket{VV}$ polarization basis. Measuring the trigger photon without resolving its polarization (which means tracing over this photon), yields 
$\rho_0=\Tr_\text{trigger}(\ket{\Psi}\!\bra{\Psi})=\tfrac{p}{2}\mbb{1}+(1-p)\ket{H}\!\bra{H}$, 
with $p=1-\cos(4\theta_p)$. To prepare the desired single-photon state, the signal photon is then sent through a half-wave plate (HWP$_1$) at angle $\theta_1$ followed by a quarter-wave plate (QWP$_1$) at angle $\phi_1$, both referenced to a horizontal fast axis. This sequence implements the transformation
\begin{equation} 
\label{eq:wp} 
\begin{split} \ket{H} \xrightarrow{\text{HWP}_1, \text{QWP}_1} & \cos{(2\theta_1)}\ket{H} + e^{i\phi_1}\sin{(2\theta_1)}\ket{V} = \ket{\psi}, 
\end{split} 
\end{equation} 
and thus, applied to $\rho_0$, it yields: 
\begin{equation} 
\rho_0 \xrightarrow{\text{HWP}_1, \text{QWP}_1} \rho = \frac{p}{2}\mbb{1} + (1 - p)\ket{\psi}\bra{\psi}. 
\end{equation} 
The state is parametrized by $\{\theta_p,\theta_1,\phi_1\}$, which relate to $\{p,\theta,\phi\}$ by identifying $\ket{H}\equiv\ket{0}$ and $\ket{V}\equiv\ket{1}$ in $\rho_\mf{c}$ and comparing $\ket{\psi}$ with $\ket{\psi_\mf{c}}$. 
This yields the simple relations $p=1-\cos(4\theta_p)$, $\theta=4\theta_1$, and $\phi=\phi_1$.

After preparing the one-qubit state $\rho_\mf{c}$, we implement nonselective measurements by entangling the signal photon with an ancilla degree of freedom (its spatial mode), which allows distinguishing the eigenstates of the measured observable. The outcome is written into the ancilla, and the measurement becomes nonselective by subsequently detecting the photon without resolving that degree of freedom.

The first step---distinguishing the eigenstates of the observable and encoding the result in the ancilla---is implemented with a PBS, which maps $H$ and $V$ polarization to distinct spatial modes and thus realizes a $\sigma_z$ measurement. 
To extend this to an arbitrary observable $A$, we rotate its eigenstates $\ket{a_i}$ to $\{\ket{H},\ket{V}\}$ using a half-wave plate followed by a quarter-wave plate, send the photon through the PBS, and then apply the inverse rotation to return to the basis of $A$. The full operation is formally described as
\begin{equation}
    \!\!\!\!\ket{a_i}\ket{0} \xrightarrow{\text{QWP}^{-1}_A,\text{HWP}^{-1}_A}  \ket{i}\ket{0} \xrightarrow{\text{PBS}} \ket{i}\ket{i} 
    \xrightarrow{\text{HWP}_A, \text{QWP}_A}
    \ket{a_i}\ket{i},
    \label{eq:ndsm}
\end{equation}
where the first qubit encodes polarization ($\ket{0}\equiv H$, $\ket{1}\equiv V$) and the second encodes path.

Finally, the nonselective map $\Phi_A$ is implemented by detecting the ancilla without recording its outcome. The procedure is performed sequentially, with one quantum non-demolition stage following another. For $\sigma_x$ and $\sigma_z$ measurements, quarter-wave plates are unnecessary, and no additional optics are required after the final PBS, since detector D1 is insensitive to polarization. The full setup is shown in Fig.~\ref{fig:setup}, and further implementation details are provided in Supplemental Material~II.

\begin{figure}[h]
    \centering
\includegraphics[width=\linewidth]{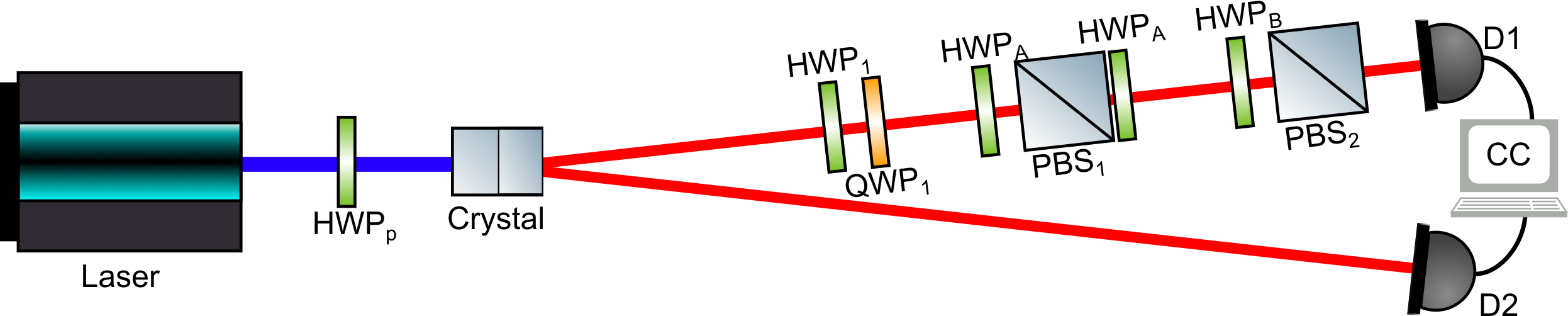}
    \caption{Experimental scheme used to obtain the photon-counting probabilities. HWP$_p$ is the element that implements the unitary modifying the parameter $p$, HWP$_1$ modifies $\theta$, and QWP$_1$ changes $\phi$. HWP$_A$ implements $\theta_A$ and, together with PBS$_1$, perform the first measurement $A$. HWP$_B$ implements $\theta_B$ and, together with PBS$_2$, performs the second nonselective measurement $B$. A final simplification involved the removal of the second HWP$_B$, which would come after PBS$_2$, as the final photon detection is performed in the polarization basis ${H, V}$, without the need for the final rotation of $\theta_B$. The elements D1 and D2 are avalanche photodetectors, and CC is the photon coincidence electronic circuit.}
    \label{fig:setup}
\end{figure}

To quantify incompatibility, we fixed $A=\sigma_x$ and $B=\sigma_z$ and performed both measurement sequences ($A$ then $B$, and $B$ then $A$). From the corresponding photon counts we obtained 
$p_\rho(b_j)$, $p_{\Phi_B(\rho)}(a_i)$, $p_\rho(a_i)$, and $p_{\Phi_A(\rho)}(b_j)$, which allowed computing $\mathcal{I}_{\mbb{C}}$ via Eq.~\eqref{I_C}. The results for qubit states aligned with the $x$-, $y$-, and $z$-axes are shown in Fig.~\ref{fig:incompatibility-xyz}.

The results match the predictions of Eq.~\eqref{angles}: compatibility occurs for maximally mixed states ($p=1$), for states orthogonal to both observables (qubit along the $y$-axis), and for states aligned with one of the measurements ($p=0$ or $p=2$), where incompatibility reaches its maximum due to noncommutativity. 
Small discrepancies between theory and experiment mainly stem from imperfect observable alignment, dominated by the $\sim1^\circ$ resolution of the waveplate rotation stages.

\begin{figure}[htb]
    \centering  \includegraphics[width=\linewidth]{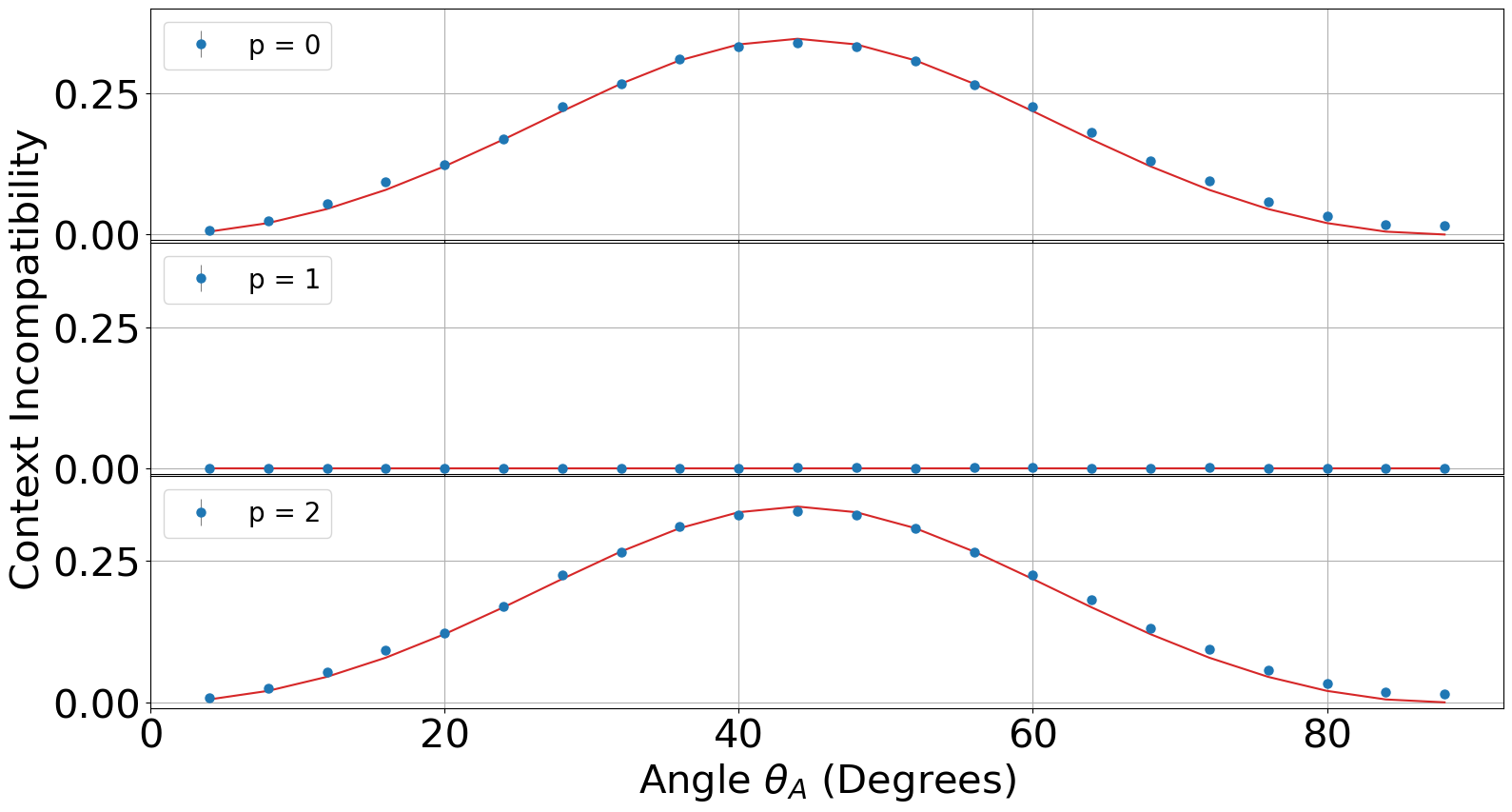}
   \caption{Theoretical prediction of the context incompatibility $\mathcal{I}_{\mbb{C}}$ (red line) and experimental results (blue dots) for the context $\mbb{C} = \{\rho_z, A, \sigma_z\}$ with fixed state interpolated at $p=0$ (top), $p=1$ (middle), and $p=2$ (bottom), as a function of the Half-Wave Plate angle $\theta_A$ that determines the observable $A$. Error bars are shown in each plot, but are not visible (on the order of $10^{-2}$). The calculations were performed using base-$e$ logarithms.}
    \label{fig:incompatibility-012}
\end{figure}

We also evaluated incompatibility for states fixed along the $z$-axis ($\theta=0$, $\phi=0$), with $B=\sigma_z$ and $A$ varied between $\sigma_z$ and $\sigma_x$. 
Here, $A=\hat{a}\cdot\vec{\sigma}$ with $\hat{a}=(\sin 2\theta_A,0,\cos 2\theta_A)$, where $\theta_A$ is the HWP$_A$ rotation angle. Results for $p=0$, $1$, and $2$ are shown in Fig.~\ref{fig:incompatibility-012}.

The results confirm that compatibility always holds for the maximally mixed state ($p=1$), independent of the observables. 
For $p=0$ or $p=2$, compatibility depends on commutation: it is observed at $\theta_A=0$ and $\pi/2$ (where $A=\sigma_z=B$ up to a sign), and incompatibility is maximal at $\theta_A=\pi/4$ (where $A=\sigma_x$). 
These results support the predicted structure of compatible and incompatible contexts and, to our knowledge, represent the first experimental quantification of context incompatibility defined independently of the quantum formalism.

\textit{Conclusions.}
We introduced a theory-independent criterion for context compatibility together with a quantitative measure of its violation. Extending previous work~\cite{MSA,Mitra_2021}, the framework is operational, order-independent, and applies beyond projective measurements and the quantum formalism. Our experimental results show, in agreement with quantum predictions, that nature does not generally satisfy context compatibility. Context incompatibility therefore emerges as a refinement of the standard notion of incompatibility underlying the Heisenberg uncertainty principle and related concepts. These results open avenues for studying context incompatibility in higher-dimensional and multipartite systems, with potential implications for quantum information processing and foundational questions such as quantum irrealism~\cite{Caetano_2024,Fucci_2024} and wave-particle duality~\cite{Dieguez_2022}.

A natural next step is to investigate the operational behavior of context incompatibility in higher dimensions, where richer state and measurement structures exist. While Eq.~\eqref{teste1} captures the quantum incompatibility bound, the full TICI quantifier remains largely unexplored beyond the qubit regime and may reveal features of incompatibility not accessible within standard quantum mechanics.

\textit{Data Availability.}
The data that support the findings of this article are openly available at \cite{storrer_2025_data}.

\textit{Acknowledgments.}
M.S. acknowledges support by the Brazilian funding agency CAPES, under the grants 88887.602269/2021-00 and 88887.883977/2023-00. R.M.A. and A.C.S.C. thank the Brazilian funding agency CNPq under Grants No. 305957/2023-6 and 308730/2023-2, respectively. S.P. acknowledges the support of the agencies FAPEMIG and CNPq (Grant No. 422300/2021-7). The authors also acknowledge support of the National Institute for the Science and Technology of Quantum Information (INCT-IQ), Grant No. 465469/2014-0.
%%%%%%%%%%%%%%%%%%%%%%%%%%%%%%%%%%%%%%%

\bibliography{ref}
\include{end_matter_v3}

\end{document}

%% file: end_matter_v3.tex
\begin{center}
\textbf{\large Supplemental Materials}
\end{center}

\section{Classical statistical mechanics example}

The motivation behind the criterion of theory-independent context compatibility, as well as its connection to elements of reality~\cite{Caetano_2024}, 
can be concretely illustrated by considering the classical statistical mechanics of a particle moving in one dimension. The system is described in phase space $q\times \pi$ by the probability density $\wp_t(q,\pi)$, satisfying the Liouville equation $\partial_t\wp_t = \{H,\wp_t\}$, where $H$ is the Hamiltonian function. A measurement of the generalized coordinate $q$ performed at time $t$, yielding outcome $\bar{q}$, leads to the transition
\begin{align}
\wp_t(q,\pi)\to \tilde{\wp}_t(q,\pi|\bar{q})=\frac{\delta(q-\bar{q})\wp_t(\bar{q},\pi)}{\int\!\!\int\! dq d\pi\,\delta(q-\bar{q})\wp_t(\bar{q},\pi)}.
\end{align}
Clearly, the description conditioned on outcome $\bar{q}$ generally differs from the original. For a nonselective measurement, one averages over all possible $\bar{q}$, weighted by their probabilities $\mf{p}_t(\bar{q}) = \int\!d\pi\,\wp_t(\bar{q}, \pi)$. It follows that
\begin{align}
\int\!d\bar{q}\,\tilde{\wp}_t(q,\pi|\bar{q})\mf{p}_t(\bar{q})=\wp_t(q,\pi).
\end{align}
The same original distribution arises from nonselective measurements of $\pi$, or from sequential measurements of $q$ and $\pi$ in any order. This shows that sequential measurements of phase-space variables do not alter the information encoded in $\wp_t(q,\pi)$. Hence, classical statistical mechanics passes test \eqref{eqsdef} with flying colors. Returning to the framework of generic theories, we emphasize the connection between our definition of context incompatibility and Fine's approach to determinism~\cite{Fine_1982}, which links this classical feature to the existence of joint probabilities. If a joint distribution $p(x_i, y_j)$ exists independently of the order in which $\ms{X}$ and $\ms{Y}$ are measured (as in classical statistical mechanics), then one has $p(x_i, y_j) = p_\ms{E}(x_i|y_j) p_\ms{E}(y_j) = p_\ms{E}(y_j|x_i) p_\ms{E}(x_i)$ (Bayes' rule). Since $\sum_j p_\ms{E}(y_j |x_i) = 1$, substituting Bayes' rule into Eq.~\eqref{M_X} yields, by marginalization, formulas \eqref{eqsdef}.

In classical statistical mechanics, this behavior reflects the epistemic nature of the state, which simultaneously encodes all measurable quantities and is not subject to any uncertainty principle arising from noncommutative algebraic structures. Consequently, measurements merely reveal pre-existing properties without inducing disturbance, and the nonselective measurement of any observable leaves the probability distribution of any other unchanged. Accordingly, all classical contexts are compatible in the sense of Eq.~\eqref{eqsdef}, provided measurements are non-invasive in the classical sense.

\section{Implementation Details of the Experiment}

Here, we provide more formal details regarding the experiment, specifically on how we implement the arbitrary qubit state $\rho_\mf{c}$ and the nonselective map \eqref{eq:nsmm}. First, we discuss the unitary transformations associated with the relevant optical elements. The unitary transformation of a half-wave plate (HWP), in the $\{\ket{H}, \ket{V}\}$ basis, is given by
\begin{equation}
    U_\text{HWP}(\theta) =
    \begin{bmatrix}
        \cos{2 \theta} & \sin{2 \theta} \\
        \sin{2 \theta} & -\cos{2 \theta}
    \end{bmatrix} \, ,
    \label{eq:hwp}
\end{equation}
where $\theta$ is the angle between the fast axis of the half-wave plate and the horizontal polarization direction. A quarter-wave plate (QWP) unitary, with its fast axis aligned horizontally, is given by
\begin{equation}
    U_\text{QWP}(\phi) =
    \begin{bmatrix}
        1 & 0 \\
        0 & e^{i\phi}
    \end{bmatrix} \, ,
    \label{eq:qwp}
\end{equation}
where $\phi$ is the relative phase shift induced between the fast and slow axes.

The polarizing beam splitter (PBS) can be understood as effectively implementing a controlled-NOT (CNOT) operation in the $\{\ket{H}, \ket{V}\} \otimes \{\ket{0}, \ket{1}\}$ basis of the Hilbert space $\mathcal{H} = \mathcal{H}_\text{pol} \otimes \mathcal{H}_\text{path}$, where the polarization acts as the control qubit and the path as the target qubit.

From Eq.~\eqref{eq:ndsm} and the matrices defined above, we can determine the angles $\theta$ and $\phi$ that implement each desired nonselective map $\Phi_A$. To do this, we focus on the final transformation in Eq.~\eqref{eq:ndsm}, as the initial transformation is its inverse, and the intermediate step is a PBS. We begin with the transformation
\begin{equation}
\label{eq:ra}
    \ket{i}_\text{pol}\ket{i}_\text{path}
    \xrightarrow{\text{HWP}_A,\text{QWP}_A}
    \ket{a_i}_\text{pol}\ket{i}_\text{path}  \, ,
\end{equation}
where it is evident that the operation acts solely on the polarization degree of freedom (the first qubit), leaving the path degree of freedom (the second qubit) unchanged.

The combined operation of a half-wave plate followed by a quarter-wave plate on the polarization computational basis $\{\ket{H},\ket{V}\}$ is described by Eqs.~\eqref{eq:hwp} and \eqref{eq:qwp}. By comparing this combined operation with each desired eigenvector $\ket{a_i}$, we can determine the required waveplate angles, and thus implement each non-destructive map $\Phi_A(\rho)$.

For our experiment, we performed nonselective measurements for observables $A \in \{\sigma_x, \sigma_z\}$. We found that the half-wave plate angles should be $\theta_x = \frac{\pi}{8}$ and $\theta_z = 0$, and the quarter-wave plate retardation should be set to $\phi = 0$ for both measurements. These results indicate that for this specific experimental realization, the quarter-wave plates are not required to perform the proposed nonselective measurement maps.

The nonselective measurement process of an observable is completed by summing over all possible outcomes associated with the path degree of freedom. This is achieved by measuring the polarization degree of freedom while disregarding the path degree of freedom, effectively performing a partial trace over the path subspace. To verify that the process in Eq.~\eqref{eq:ndsm}, after tracing out the ancilla, reproduces the expected result, we proceed as follows.

Let $R_A$ be the unitary operation on the polarization qubit that implements the evolution described in Eq.~\eqref{eq:ra}. We apply the sequence of operations in Eq.~\eqref{eq:ndsm} to an arbitrary initial polarization state $\rho$ and the path qubit initial state $\ket{0}$:
\begin{equation}
    \begin{split}
    \rho \otimes \ket{0}\bra{0} & \xrightarrow{R^{\dagger}_A \otimes I} R^{\dagger}_A \rho R_A \otimes \ket{0}\bra{0} \\
        &\xrightarrow{\text{PBS}} \ket{0}\bra{0}R^{\dagger}_A \rho R_A\ket{0}\bra{0} \otimes \ket{0}\bra{0} \\ 
        &+ \ket{1}\bra{1}R^{\dagger}_A \rho R_A\ket{1}\bra{1} \otimes \ket{1}\bra{1} \\
        &+ \ket{0}\bra{0}R^{\dagger}_A \rho R_A\ket{1}\bra{1} \otimes \ket{0}\bra{1} \\
        &+ \ket{1}\bra{1}R^{\dagger}_A \rho R_A\ket{0}\bra{0} \otimes \ket{1}\bra{0} \\
        &\xrightarrow{R_A \otimes I} R_A \ket{0}\bra{0}R^{\dagger}_A \rho R_A\ket{0}\bra{0}R^{\dagger}_A \otimes \ket{0}\bra{0} \\ 
        &+ R_A \ket{1}\bra{1}R^{\dagger}_A \rho R_A\ket{1}\bra{1}R^{\dagger}_A \otimes \ket{1}\bra{1} \\
        &+ R_A \ket{0}\bra{0}R^{\dagger}_A \rho R_A\ket{1}\bra{1}R^{\dagger}_A \otimes \ket{0}\bra{1} \\
        &+ R_A \ket{1}\bra{1}R^{\dagger}_A \rho R_A\ket{0}\bra{0}R^{\dagger}_A \otimes \ket{1}\bra{0} \\
        &= \ket{a_0}\bra{a_0} \rho \ket{a_0}\bra{a_0} \otimes \ket{0}\bra{0} \\ 
        &+ \ket{a_1}\bra{a_1} \rho \ket{a_1}\bra{a_1} \otimes \ket{1}\bra{1} \\
        &+ \ket{a_0}\bra{a_0} \rho \ket{a_1}\bra{a_1} \otimes \ket{0}\bra{1} \\
        &+ \ket{a_1}\bra{a_1} \rho \ket{a_0}\bra{a_0} \otimes \ket{1}\bra{0} \\
        &\equiv \rho_\text{res}, 
    \end{split}
\end{equation}
where we used that $R^{-1}_A  =R^{\dagger}_A$ and $R_A\ket{i} = \ket{a_i}$.  Ultimately, tracing out the path degree of freedom, we find
\begin{equation}
    \begin{split}
        \rho_\text{final} &= \Tr_\text{path}\{\rho_\text{res}\} \\
        &= \ket{a_0}\bra{a_0} \rho \ket{a_0}\bra{a_0} + \ket{a_1}\bra{a_1} \rho \ket{a_1}\bra{a_1},
    \end{split}
\end{equation}
which is identical to Eq.~\eqref{eq:nsmm} with Kraus operators $A_i = \ket{a_i}\bra{a_i}$, as desired.